\documentclass{article}
\usepackage{spconf,amsmath,graphicx,hyperref}
\usepackage{cite}
\usepackage{amsmath,amssymb,amsfonts}
\usepackage{algorithmic}
\usepackage{graphicx}
\usepackage{textcomp}
\usepackage{xcolor}
\usepackage{lipsum}
\usepackage{multirow}
\usepackage[table,xcdraw]{xcolor}
\usepackage{booktabs}

\title{Dual Data Scaling for Robust Two-Stage User-Defined Keyword Spotting}
\name{
    Zhiqi Ai\textsuperscript{1},
    Han Cheng\textsuperscript{1},
    Yuxin Wang\textsuperscript{1},
    Shiyi Mu\textsuperscript{1},
    Shugong Xu\textsuperscript{2,*},
    Yongjin Zhou\textsuperscript{1,*}\thanks{\textsuperscript{*}Corresponding authors}
}

\address{
\textsuperscript{1}Shanghai University, Shanghai, China \\
\textsuperscript{2}Xi’an Jiaotong Liverpool University, Suzhou, China
}
\begin{document}
%
\maketitle
\begin{abstract}
In this paper, we propose DS-KWS, a two-stage framework for robust user-defined keyword spotting. It combines a CTC-based method with a streaming phoneme search module to locate candidate segments, followed by a QbyT-based method with a phoneme matcher module for verification at both the phoneme and utterance levels. To further improve performance, we introduce a dual data scaling strategy: (1) expanding the ASR corpus from 460 to 1,460 hours to strengthen the acoustic model; and (2) leveraging over 155k anchor classes to train the phoneme matcher, significantly enhancing the distinction of confusable words. Experiments on LibriPhrase show that DS-KWS significantly outperforms existing methods, achieving 6.13\% EER and 97.85\% AUC on the Hard subset. On Hey-Snips, it achieves zero-shot performance comparable to full-shot trained models, reaching 99.13\% recall at one false alarm per hour.
\end{abstract}
\begin{keywords}
user-defined keyword spotting, two-stage, data scaling, wake-word detection
\end{keywords}
\section{Introduction}
\label{sec:intro}

The widespread use of smart devices and conversational terminals (e.g., “XiaoZhi AI”) has created extensive application scenarios for user-defined keyword spotting (UDKWS) \cite{ baseline_cmcd, baseline_adakws, ai24_interspeech, baseline_phonmatchnet} to meet users’ personalized needs. However, compared with predefined keywords trained on large-scale target-word datasets \cite{wekws}, user-defined keywords still exhibit a performance gap, as they primarily rely on zero-shot capabilities \cite{ai24_interspeech, kim25d_interspeech, baseline_adakws}.

Traditional ASR-based methods are often costly and perform poorly on out-of-vocabulary words. Recently, lightweight phoneme-level methods have shown strong performance on keyword spotting tasks when trained on sufficiently large ASR corpora \cite{cdc_kws, tdt_kw, mfa-kws, ntc, u2kws,yang22n_interspeech}. \cite{tdt_kw} extends the standard RNN-T architecture, enabling it to skip input frames based on predicted durations, which significantly accelerates inference. \cite{cdc_kws} proposes a CTC streaming decoding algorithm that achieves higher recall than ASR- and graph-based baselines. \cite{mfa-kws, u2kws, yang22n_interspeech} adopt multi-stage frameworks to achieve more precise keyword detection; although these models are considerably more lightweight than traditional ASR systems, they still require substantial fine-tuning data to ensure optimal performance in practical deployment.

Lightweight query-by-example (QbyE) methods compare registered features with query audio, enabling low-power, low-latency UDKWS \cite{wang21ea_interspeech, kim2019query}. Some studies use registered audio templates for coarse-grained matching (e.g., similarity-based), known as QbyA \cite{wang21ea_interspeech, kirandevraj2022generalized}. In contrast, \cite{baseline_cmcd, baseline_emkws, baseline_ced, baseline_adakws, cacd, SLiCK} leverage text-based registration (QbyT) for more stable and superior performance, while \cite{ai24_interspeech, plcl} adopt a multimodal approach (text\&audio, QbyAT), achieving state-of-the-art results on the LibriPhrase dataset. Additionally, \cite{baseline_ced, baseline_adakws, ai24_interspeech, plcl} show that QbyE methods are particularly effective at distinguishing confusable words. However, these methods are trained on pre-segmented pairs, and rely on sliding-window detection at inference, which can produce inaccurate boundaries and more false alarms.

To address these challenges, we propose DS-KWS, a two-stage framework for robust user-defined keyword spotting. In the first stage, a CTC-based branch is combined with a streaming phoneme search module to detect candidate segments, whereas the second stage uses a QbyT-based phoneme matcher to verify them at both phoneme and utterance levels. We introduce a dual data scaling strategy:(1) expanding the ASR training corpus to 1,460 hours by combining LibriSpeech-460 with GigaSpeech-1000 to strengthen the acoustic model; and (2) expanding the LibriPhrase-460 training set with GigaPhrase-1000, which extends it to 155k anchor classes and substantially increases both the number of anchor phrases and the coverage of confusable words. Experiments show that DS-KWS significantly outperforms previous methods on LibriPhrase, achieving 6.13\% EER and 97.85\% AUC on LibriPhrase-Hard. On Hey-Snips, it achieves zero-shot performance comparable to full-shot trained models, reaching 99.13\% recall at one false alarm per hour. 

\section{Proposed Method}

\begin{figure*}[htbp]
\centerline{\includegraphics[width=0.9\linewidth]{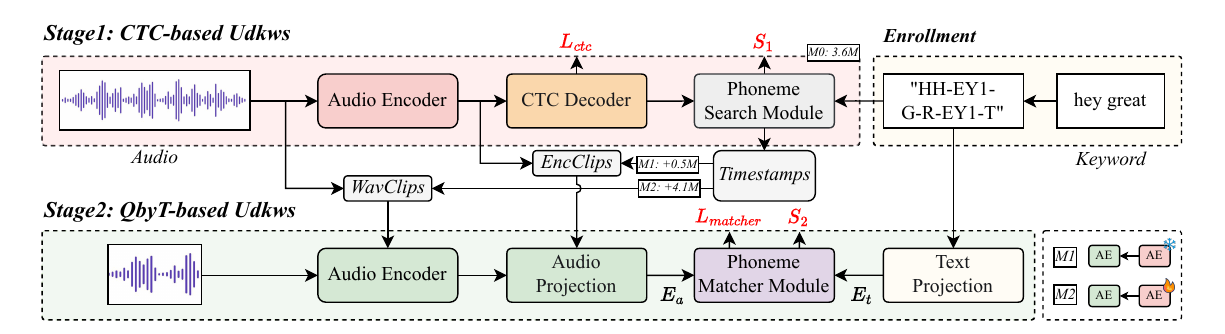}}
\caption{Overall architecture of the DS-KWS model. The CTC branch extracts phoneme sequences, and the phoneme search module outputs score $S_1$ and candidate segments. The phoneme matcher produces the second-stage score $S_2$.}
\label{fig.Overall}
\end{figure*}

\begin{figure}[htbp]
\centerline{\includegraphics[width=0.85\linewidth]{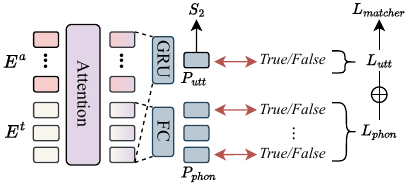}}
\caption{Implementation of the Phoneme Matcher Module.}
\label{fig.pm}
\end{figure}

In this section, we present the proposed DS-KWS. The overall architecture is shown in Figure~\ref{fig.Overall}. The CTC-based branch consists of an audio encoder, a CTC decoder, and a target phoneme search module, while the QbyT branch includes an audio encoder (frozen or fine-tuned), two modality projection layers, and a phoneme matcher (Figure~\ref{fig.pm}).
\subsection{CTC-Based Branch}

\textbf{Feature Extractor:} Inspired by \cite{baseline_ced, baseline_emkws, ai24_interspeech, kim25d_interspeech}, we adopt the Conformer architecture \cite{conformer} as the audio encoder, transforming the input speech into a sequence of audio embeddings. Let the input speech be denoted as $F_a$, which is fed into the audio encoder to produce $E_a$.
\begin{equation}
    E_a = (e_1,e_2,\dots,e_T) 
\end{equation}
where $T$ is the length of the encoder output.

For the keyword text input, the CTC target is represented as a sequence of 71 phonemes (including a \textit{blank} token) obtained via G2P \cite{ai24_interspeech, baseline_cmcd, mfa-kws}, and then converted into token indices using a character-level tokenizer:
\begin{equation}
    X = \textbf{Tokenizer}(\textbf{G2P}(text)) \in \mathbb{Z}^{71} 
\end{equation}

The audio embeddings $E_a$ are trained with CTC Loss and mapped by the CTC layer to a token probability distribution.
\begin{equation}
    c = (c_1,c_2,\dots,c_T) 
\end{equation}
where $c_t$ denotes the predicted probability distribution over the 71 phoneme tokens at time step $t$.

\textbf{Phoneme Search Module:} The phoneme search module is a streaming CTC decoding algorithm proposed in \cite{cdc_kws, mfa-kws} , which retrieves candidate paths corresponding to the target phoneme sequence from the CTC output. To handle fuzzy targets (e.g., \textit{Hi Snips} vs. \textit{Hey Snips}, differing only in AY1/EY1), we merge the probabilities of fuzzy phonemes. Formally, for a phoneme $x_t$ with fuzzy set $\mathcal{F}(x_t)$, we define:
\begin{equation}
S_\text{agg}(x_t) = \sum_{f \in \mathcal{F}(x_t)} S(f),
\end{equation}
where $S(f)$ denotes the CTC probability of phoneme $f$. This aggregation improves recall by accepting variations of the target keyword.

\subsection{Query-by-Text (QbyT) Branch}
\label{sec:m0}
\textbf{Feature Extractor:} In the second stage, candidate segments from the CTC branch are verified using a QbyT-based phoneme matcher trained on 155k anchor classes to learn cross-modal alignment. Two modes are considered (Figure~\ref{fig.Overall}).\textbf{M1:} Candidate features are directly cropped from the Stage~1 encoder outputs according to the \textit{timestamps}, and then a lightweight projection layer is applied for dimensional mapping. This adds only $\sim$0.5M parameters. 
\textbf{M2:} Candidate segments are extracted as raw speech clips using the \textit{timestamps} and re-encoded by a trainable audio encoder in Stage 2. This produces more discriminative representations, adding $\sim$3.6M parameters compared to M1.

\textbf{Lightweight registration module:} Previous works typically rely on large pre-trained encoders for anchor registration, such as DistilBERT \cite{ai24_interspeech,baseline_emkws}, G2P-based embeddings \cite{ai24_interspeech, baseline_ced}, and audio-text registration modules\cite{plcl, ai24_interspeech}. While these approaches achieve strong performance, they incur substantial offline computational overhead. In contrast, our method uses a simple \textbf{nn.Embedding} that directly maps phoneme indices into the feature space as $E_t$, drastically reducing both computational and memory requirements for registration.

\begin{table}[htbp]
\caption{Experimental results of the proposed DS-KWS model on the Libriphrase dataset compared with the baseline.}
\centering
\resizebox{\linewidth}{!}{
\begin{tabular}{@{}lccccc@{}}
\toprule
\multirow{2}{*}{\textbf{Method}} & \multirow{2}{*}{\textbf{\# Params}} & \multicolumn{2}{c}{\textbf{AUC (\%) ↑}} & \multicolumn{2}{c}{\textbf{EER (\%) ↓}} \\ \cmidrule(l){3-6} 
             &      & \textbf{$\textbf{LP}_{\textbf{H}}$} & \textbf{$\textbf{LP}_{\textbf{E}}$} & \textbf{$\textbf{LP}_{\textbf{H}}$} & \textbf{$\textbf{LP}_{\textbf{E}}$} \\ \midrule
CMCD \cite{baseline_cmcd}         & 0.7M & 73.58              & 96.70              & 32.90              & 8.42               \\
EMKWS \cite{baseline_emkws}        & 3.7M & 84.21              & 97.83              & 23.36              & 7.36               \\
CED \cite{baseline_ced}         & 3.6M & 92.70              & 99.84              & 14.40              & 1.70               \\
SLiCK \cite{SLiCK} & 0.6M & 94.90              & 99.82              & 11.10              & 1.78               \\
AdaKWS-Small \cite{baseline_adakws} & 109M & 95.09              & 99.82              & 11.48              & 1.21               \\
MM-KWS-T \cite{ai24_interspeech}    & 3.9M & 95.36              & 99.94              & 10.41              & 0.82               \\
PLCL-T \cite{plcl}      & 40M  & 95.56              & 99.95              & 9.96               & 1.21               \\
W-CTC \cite{kim25d_interspeech}        & 3.6M & 95.93              & 99.95              & 10.21              & 0.91               \\ \midrule
MM-KWS-AT \cite{ai24_interspeech}   & 3.9M & 96.25              & 99.95              & 9.30               & 0.68               \\
PLCL-AT \cite{plcl}     & 40M  & 96.59              & 99.97              & 8.47               & 0.57               \\ \midrule
DS-KWS-M1    & 4.1M & 95.77              & 99.98              & 10.02              & 0.52               \\
DS-KWS-M2    & 4.1M & \textbf{97.85}     & \textbf{99.98}     & \textbf{6.13}      & \textbf{0.45}      \\ \bottomrule
\end{tabular}
}
\label{table:lp}
\end{table}

\textbf{Phoneme Matcher Module:} As shown in Figure~\ref{fig.pm}, our phoneme matcher module uses a lightweight attention mechanism to align audio features ($E_a$) with the phoneme sequence ($E_t$). These features are then processed by a discriminator for phoneme- and utterance-level matching. The resulting utterance-level score $P_{utt}$ serves as the output score ($S2$). The module is trained using a joint loss $\mathcal{L}_\text{matcher}$:
\begin{equation}
\mathcal{L}_\text{matcher} = \mathcal{L}_\text{utt} + \mathcal{L}_\text{phon}
\end{equation}

where $\mathcal{L}_\text{utt}$ and $\mathcal{L}_\text{phon}$ denote the utterance- and phoneme-level matching losses, respectively. 

The overall training objective of our framework can be expressed as:
\begin{equation}
\mathcal{L}_\text{total} = \mathcal{L}_\text{CTC} + \mathcal{L}_\text{matcher}
\end{equation}

\section{Experiments}

\subsection{Datasets}

\textbf{ASR Training:} We use LibriSpeech-100 (\textbf{LS-100}) and LibriSpeech-460 (\textbf{LS-460}), two “clean” training subsets of LibriSpeech containing 100 and 460 hours of audio, respectively. To further expand the training data, we include GigaSpeech-1000 (\textbf{GS-1000}), derived from the “Middle” portion of GigaSpeech. These datasets are combined to form the \textbf{LS-GS-1460} training set.

\textbf{Phrase Training:} We constructed LibriPhrase (LP) following \cite{baseline_cmcd, baseline_emkws, baseline_ced, kim25d_interspeech, baseline_adakws, ai24_interspeech, plcl}. Only anchors with durations between 0.5–2 seconds and at least 10 instances were retained. The training data includes \textbf{LP-100} ($\sim$12k classes) and \textbf{LP-460} ($\sim$78k classes). To study data scaling for the second stage, we randomly sampled two subsets from LP-460: \textbf{LP-460 (R2W)} and \textbf{LP-460 (R4W)}, containing 20k and 40k classes, respectively. Finally, we combined LS-460 with GigaPhrase-1000 (\textbf{GP-1000}) to construct the \textbf{LS-GP-1460} training set, containing $\sim$155k classes.
   
\textbf{Evaluations:} First, for assessing ASR model performance, we use the LibriSpeech clean and other test sets ($\textbf{LS}_\textbf{clean}$ and $\textbf{LS}_\textbf{other}$). Second, for keyword spotting, we use LibriPhrase, derived from LibriSpeech-other-500, serving as the main benchmark. The test set is further divided into LibriPhrase Easy ($\textbf{LP}_{\textbf{E}}$) and LibriPhrase Hard ($\textbf{LP}_{\textbf{H}}$) subsets. Evaluation metrics primarily include Equal Error Rate (EER) and Area Under the ROC Curve (AUC). Third, we use the \textbf{Hey-Snips} dataset, where only the test set is used, containing 2,529 positive and 20,543 negative utterances. Evaluation is conducted using Recall@FAR, which measures recall at a fixed false alarm rate.

\begin{table}[]
\caption{Dual Data Scaling of the proposed DS-KWS.}
\centering
\resizebox{0.9\linewidth}{!}{
\begin{tabular}{@{}lcccccc@{}}
\toprule
\multirow{2}{*}{\textbf{Setting}} & \multicolumn{2}{c}{\textbf{P-WER (\%) ↓}}        & \multicolumn{2}{c}{\textbf{AUC (\%) ↑}} & \multicolumn{2}{c}{\textbf{EER (\%) ↓}} \\ \cmidrule(l){2-7} 
                                  & \textbf{$\textbf{LS}_{\textbf{clean}}$} & \textbf{$\textbf{LS}_{\textbf{other}}$} & \textbf{$\textbf{LP}_{\textbf{H}}$}     & \textbf{$\textbf{LP}_{\textbf{E}}$}    & \textbf{$\textbf{LP}_{\textbf{H}}$}     & \textbf{$\textbf{LP}_{\textbf{E}}$}    \\ \midrule
\multicolumn{7}{l}{\textit{Stage1: LS-100}}                 \\ \midrule
LP-100-M1     & 6.98 & 18.79 & 91.78 & 99.85 & 15.34 & 1.35 \\
LP-100-M2     & -    & -     & 93.10 & 99.88 & 13.71 & 1.14 \\ \midrule
\multicolumn{7}{l}{\textit{Stage1: LS-460}}                 \\ \midrule
LP-460-M1     & \textbf{4.44} & \underline{13.39} & 95.33 & 99.96 & 10.78 & 0.72 \\
LP-460-M2     & -    & -     & \underline{97.03} & \underline{99.96} & \underline{7.97}  & \underline{0.59} \\ \midrule
\multicolumn{7}{l}{\textit{Stage1: LS-GS-1460}}             \\ \midrule
LP-GP-1460-M1 & \underline{4.45} & \textbf{11.80} & 95.77 & 99.98 & 10.02 & 0.52 \\
LP-GP-1460-M2                     & -                   & -                   & \textbf{97.85}  & \textbf{99.98} & \textbf{6.13}   & \textbf{0.45}  \\ \bottomrule
\end{tabular}
}
\label{tab:dual_data_scaling}
\end{table}

\subsection{Training Details}
\textbf{Audio Encoder:} The audio encoder uses 80-channel mel spectrograms (25 ms window, 10 ms hop) as input. The encoder architecture is a 6-layer Conformer with encoder dimension 144, linear dimension 576, convolution kernel size 3, and 4 attention heads, with $\sim$3.6M trainable parameters.

\textbf{Training:} The first stage trains the audio encoder using WeNet with CTC loss. The phoneme vocabulary consists of 71 symbols (including the blank token) generated via G2P. Separate models are trained on 100h (LS-100), 460h (LS-460), and 1,460h (LS-GS-1460) of audio, with $\sim$3.6M trainable parameters. The second stage employs a lightweight QbyT branch, comprising 2 Transformer layers, a GRU, and several fully-connected layers. Text features are encoded via \textbf{nn.Embedding}, with the total number of parameters $\sim$0.5M. 

\textbf{Dual Data Scaling:} To systematically study the effect of training data size on model performance, we adopt a dual-stage strategy. In Stage 1, the CTC-based audio encoder is trained on progressively larger datasets (LS-100, LS-460, and LS-GS-1460) to assess how acoustic model capacity scales with data. In Stage 2, the phoneme matcher (QbyT branch) is trained on corresponding phrase-level datasets with increasing anchor class diversity: LP-100 ($\sim$12k classes), LP-460 ($\sim$78k classes), and LP-GP-1460 ($\sim$155k classes).

\section{Results}
\subsection{Comparative Evaluation of DS-KWS}

Table \ref{table:lp} presents a comparative evaluation of the proposed DS-KWS model against recent methods on the LibriPhrase dataset. The results demonstrate that DS-KWS achieves significant performance improvements, particularly on the $\text{LP}_\text{H}$ subset. Notably, DS-KWS-M2 attains an AUC of 97.85\% and an EER of 6.13\% on $\text{LP}_\text{H}$, outperforming the ASR-pretrained W-CTC \cite{kim25d_interspeech} and substantially surpassing multimodal registration approaches such as MM-KWS-AT \cite{ai24_interspeech} and PLCL-AT \cite{plcl}. On the $\text{LP}_\text{E}$ subset, DS-KWS-M2 also excels, achieving an  EER of 0.45\%, the best among all compared methods, highlighting its effectiveness across both easy and hard subsets. Furthermore, even without data augmentation using GS-1000 and GP-1000, LP-460-M2 achieves an EER of 7.97\% (Table \ref{tab:dual_data_scaling}), further validating the effectiveness of the dual-stage training strategy. Importantly, under the M1 mode with a frozen encoder, DS-KWS still delivers competitive results ($\text{LP}_\text{H}$ EER 10.02\%), demonstrating the robustness of the framework even when encoder parameters are fixed.

\subsection{Dual Data Scaling Evaluation of DS-KWS}
\textbf{Scaling the Training Corpus:} We evaluate the impact of dual data scaling on DS-KWS by varying both the training corpus size and the encoder configuration. As shown in Table~\ref{tab:dual_data_scaling}, in Stage 1 training, increasing the dataset size leads to a significant reduction in P-WER, particularly on the $\text{LS}_\text{others}$ set (from 18.79\% to 11.80\%). Additionally, under the M1 model, LP-GP-1460-M1 shows notable improvement over LP-100-M1. In the M2 mode, due to the simultaneous expansion of pretraining and fine-tuning data, LP-460-M2 and LP-GP-1460-M2 achieve substantial performance gains, reaching EER of 7.97\% and 6.13\%, respectively, on the $\text{LP}_\text{H}$.

\textbf{Effect of Anchor Scaling in Stage 2:} To further investigate the impact of data scale in stage 2, we conducted an ablation study by varying the number of anchor classes used during training, with results summarized in Table~\ref{tab:stage2_ablation}. The experiments show that as the anchor set progressively expands from 12k → 20k → 40k → 78k → 155k, the model’s performance steadily improves. Specifically, the EER on the $\text{LP}_\text{H}$ decreases from 13.38\% to 10.65\%, while the AUC increases from 93.22\% to 95.45\%. These results indicate that increasing the number of anchor classes can significantly enhance the model’s ability to discriminate between confusable keywords.

\begin{table}[]
\label{table:scaling2}
\caption{Effect of Anchor Scaling for DS-KWS. The audio encoder, \textbf{pretrained on LS-460}, is frozen during training.}
\centering
\resizebox{\linewidth}{!}{
\begin{tabular}{@{}lccccc@{}}
\toprule
\multirow{2}{*}{\textbf{Setting}} & \multirow{2}{*}{\textbf{\# Anchors}} & \multicolumn{2}{c}{\textbf{AUC (\%) ↑}} & \multicolumn{2}{c}{\textbf{EER (\%) ↓}} \\ \cmidrule(l){3-6} 
             &      & \textbf{$\textbf{LP}_{\textbf{H}}$} & \textbf{$\textbf{LP}_{\textbf{E}}$} & \textbf{$\textbf{LP}_{\textbf{H}}$} & \textbf{$\textbf{LP}_{\textbf{E}}$} \\ \midrule
LP-100       & 12k  & 93.22       & 99.88       & 13.38       & 1.19        \\
LP-460 (R2W) & 20k  & 93.95       & 99.94       & 12.50       & 0.82        \\
LP-460 (R4W) & 40k  & 94.75       & 99.96       & 11.62       & 0.69        \\
LP-460       & 78k  & 95.33       & 99.96       & 10.78       & 0.72        \\
LP-GP-1460   & 155k & \textbf{95.45}       & \textbf{99.97}       & \textbf{10.65}       & \textbf{0.64}        \\ \bottomrule
\end{tabular}
}
\label{tab:stage2_ablation}
\end{table}

\begin{table}[]
\caption{Zero-shot performance of the proposed DS-KWS.}
\centering
\resizebox{0.9\linewidth}{!}{
\begin{tabular}{@{}lccc@{}}
\toprule
\multirow{2}{*}{\textbf{Method}} & \multirow{2}{*}{\textbf{Training Data}} & \multicolumn{2}{c}{\textbf{Recall (\%) @FARs}} \\ \cmidrule(l){3-4} 
                                 &                                         & \textbf{0.5}           & \textbf{1}          \\ \midrule
RIL-KWS \cite{zhang20u_interspeech}                          & \multirow{2}{*}{Official Snips}         & 96.47                  & 97.18               \\
MDTC \cite{wekws}                             &                                         & 99.88                  & 99.92               \\ \midrule
DS-KWS-M0 (3.6M)                        & \multirow{3}{*}{Zero-Shot}              & 98.89                  & 98.97               \\
DS-KWS-M1 (4.1M)                       &                                         & 98.58                  & 98.93               \\
DS-KWS-M2 (7.7M)                       &                                         & 98.97                  & 99.13               \\ \bottomrule
\end{tabular}
}
\label{tab:snips_results}
\end{table}
\begin{figure}[htbp]
\centerline{\includegraphics[width=\linewidth]{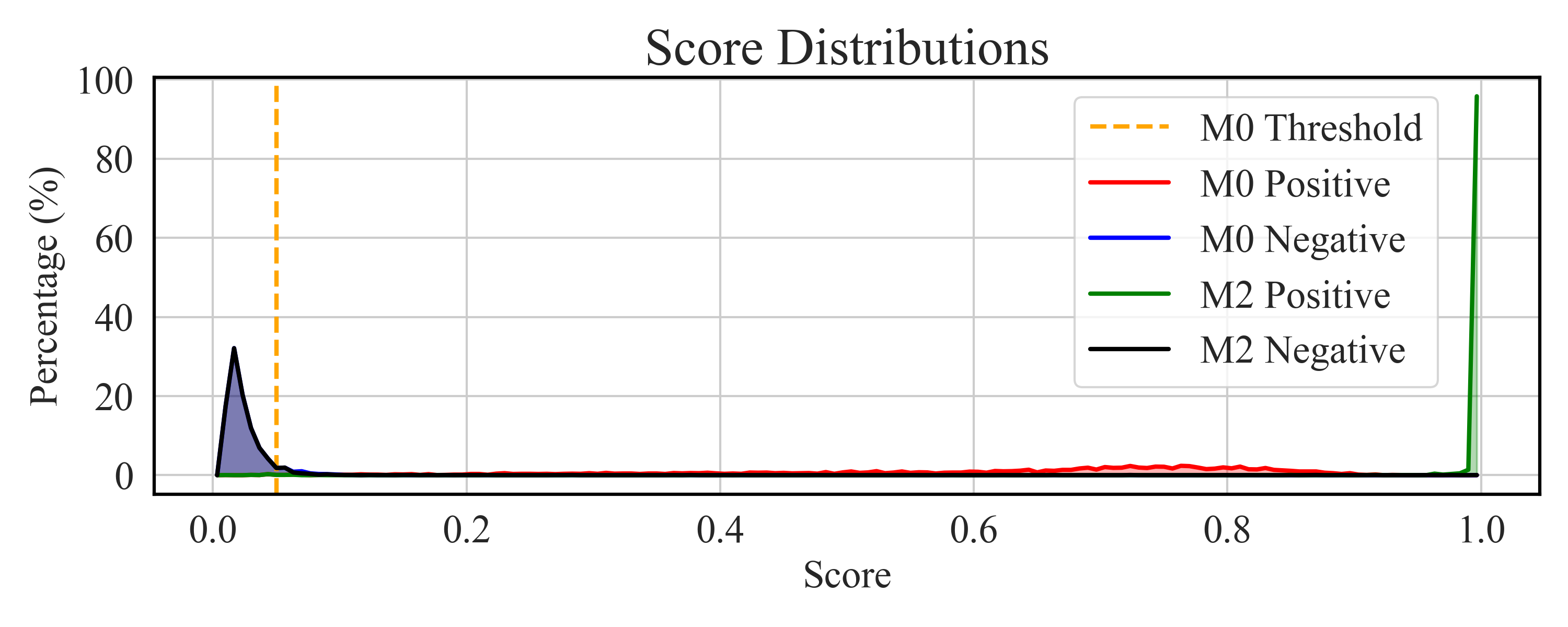}}
\caption{Comparison of score distributions for DS-KWS-M0 and -M2.}
\label{fig.wdd}
\end{figure}

\subsection{Zero-shot Performance in Wake-word Detection}
We evaluated the zero-shot capabilities of DS-KWS on the Snips dataset (Table~\ref{tab:snips_results}). In M1 mode, with a frozen encoder, the 0.5M-parameter phoneme matcher exhibits limited generalization, performing slightly worse than using CTC alone. Overall, however, DS-KWS demonstrates strong recognition of unseen wake-up words, with DS-KWS-M2 achieving recalls of 98.97\% at FAR = 0.5 and 99.13\% at FAR = 1, approaching the performance of full-shot training. Score distribution visualizations (Figure~\ref{fig.wdd}) indicate that while the Stage-1 CTC outputs separate positive and negative samples, the positive scores are widely distributed (0.04–1). After preliminary filtering in Stage 1, Stage-2 scores clearly distinguish positives from negatives, highlighting the strong zero-shot generalization of DS-KWS-M2.

\section{Conclusions}
In this paper, we propose DS-KWS, a two-stage framework for robust user-defined keyword spotting. The first stage integrates a CTC-based branch with a streaming phoneme search module, while the second stage employs a QbyT-based phoneme matcher to verify candidate segments at both the phoneme and utterance levels. By leveraging a dual data scaling strategy, DS-KWS effectively distinguishes confusable keywords and generalizes to unseen user-defined phrases. In future work, we plan to explore fine-tuning strategies for a broader set of wake-up words in real-world scenarios and investigate deployment on low-resource platforms to validate its practical effectiveness.

\bibliographystyle{IEEEbib}
\bibliography{strings,refs}

\end{document}